\documentclass{appolb}
\usepackage{epsfig}
\usepackage{cite}

\newcommand{\psb}{\bar{\psi}}

\newcommand{\nn}{\nonumber}
\begin{document}
\pagestyle{plain}
\eqsec  
\title{\bf MULTI-PARTICLE PROCESSES  IN THE STANDARD \\\\
\vspace{0.8cm}
MODEL   WITHOUT FEYNMAN DIAGRAMS 
\thanks{Presented at the  XXIX International Conference
of Theoretical Physics, {\it ``Matter to the Deepest''},
Recent Developments in Physics of Fundamental Interactions,
Ustron, Poland, 8 - 14 September 2005.} ~\thanks{{\tt IFJPAN-V-2005-12}}}
\author{Costas G. Papadopoulos
\address{Institute of Nuclear Physics,
        NCSR Demokritos\\ 15-310 Athens, Greece}
\and
Ma\l gorzata  Worek
\address{ 
Institute of Nuclear Physics,
        NCSR Demokritos\\ 15-310 Athens, Greece\\
Institute of Nuclear
        Physics, PAS\\ Radzikowskiego 152, 31-3420  Cracow,
        Poland}
}
\maketitle

\begin{abstract}

A method to efficiently compute, in a automatic way, helicity amplitudes for
arbitrary  scattering processes at leading order 
in the Standard Model is presented.
The scattering amplitude is evaluated recursively
through a set of Dyson-Schwinger equations. The computational cost of this
algorithm grows asymptotically as $3^n$, where $n$ is the number of particles
involved in the process, compared to $n!$ in the
traditional Feynman graphs approach. Unitary gauge is used and mass effects are
available as well. Additionally,
the color and helicity structures are appropriately transformed so the usual
summation is replaced by Monte Carlo techniques.
Some results related to the production of vector bosons and the Higgs boson 
in association 
with jets are also presented.
\end{abstract}

\section{Introduction}

\begin{figure}[!ht]
\begin{center}
\epsfig{file=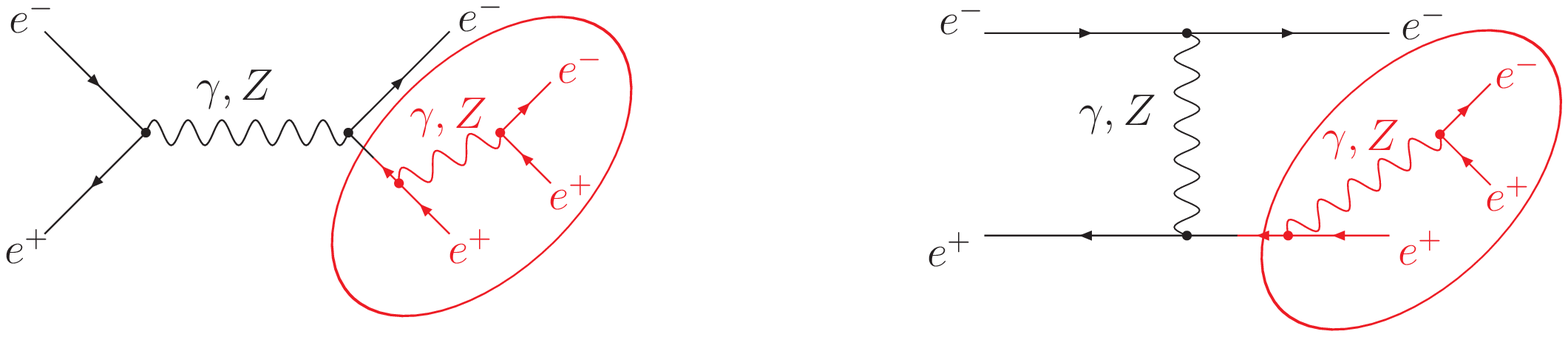,width=120mm,height=30mm}
\end{center}
\caption
{An example of common parts of the amplitudes (in red) for the 
$e^{+}e^{-}\rightarrow e^{+}e^{-}e^{+}e^{-}$ process.}
\label{diagram}
\end{figure}

Multi-particle and multi-jet final states are of great importance at the
TeVatron and at the future  LHC or $e^{+}e^{-}$ Linear Collider. 
They serve both as signals and as important backgrounds
to many new and already discovered physics channels. 
As an example the production and decay of top 
quarks, Higgs boson(s) or SUSY particles can be mentioned.  
A typical background is the production of weak vector bosons 
in association with jets. Among others the 
proper evaluation of the eight jet QCD background will be needed.  
To describe the production process of a number of particles the 
corresponding amplitudes have to be constructed. This usually results 
in a very large number of terms, such that their automated construction 
and evaluation becomes the only solution. Apart from handling the number 
of amplitudes, which grows factorially with the number of particles, the
 integration over the multidimensional phase space of the final state 
particles represents a formidable task.   In the past years various
solutions to deal with these problems, implemented as different codes, 
have been presented. Either they are based on traditional methods of 
constructing Feynman diagrams or alternative methods with recursive 
equations are implemented \cite{Berends:1987cv,Berends:1987me,Mangano:1987xk,Mangano:1987kp,Berends:1988yn,Mangano:1988kk,Berends:1989hf,Berends:1990ax,Mangano:1990by,Caravaglios:1995cd,Draggiotis:1998gr,Caravaglios:1998yr,Draggiotis:2002hm}. 
The new formalism based on the Dyson-Schwinger 
equations recursively defines one-particle off-shell Green function.
It does not involve any
calculation of individual diagrams but various off-shell subamplitudes are 
regroupped in such a way that as little of the computation as possible is 
repeated. On the contrary, in the traditional approach, the 
same parts of different Feynman diagrams are recalculated all over again,
see Fig.\ref{diagram}, increasing the number of steps that should be done in 
order to get the full amplitude. The recursive 
approach significantly decreases the factorial growth of the number of terms 
to be calculated with the number of particles down to $4^{n}$ or 
$3^{n}$ \footnote{To reduce the computational complexity down to an 
asymptotic $3^n$, each four-boson vertex must be replaced
with a three-boson vertex \eg by introducing an auxiliary field
represented by the antisymmetric tensor $H^{\mu\nu}$, see 
\cite{Draggiotis:2002hm} for details.}.

Some examples of automatic parton level generators for any processes in
the Standard  Model are  \eg {\tt CompHEP} 
\cite{Boos:1994xb,Pukhov:1999gg,Boos:2004kh}, 
{\tt MadGraph/MadEvent}\cite{Stelzer:1994ta,Maltoni:2002qb},  
{\tt AMEGIC++} \cite{Krauss:2001iv} and the {\tt HELAC/PHEGAS}
 package \cite{Kanaki:2000ey,Kanaki:2000ms}.
Codes designed for specific processes are \eg
{\tt GRACE} \cite{Kaneko:1991ym,Ishikawa:1993qr} as well as 
 {\tt Alpgen} \cite{Mangano:2002ea}. 
Very recently also 
on shell recursive equations have been 
proposed \cite{Britto:2004ap,Britto:2005fq}. However,  event 
generators based on this new method are not publicly available yet.

In this article the algorithm based on  Dyson-Schwinger
recursive equations is briefly reviewed. It has been 
implemented  as a new version 
of the multipurpose Monte Carlo generator  {\tt HELAC}  
in order to  efficiently obtain
cross sections for arbitrary  multi-particle and multi-jet processes in the
Standard Model.
 
\section{Dyson-Schwinger Recursive Equations}

Dyson-Schwinger equations give recursively the $n-$point Green's functions
in terms of the $1-$, $2-$,$\ldots$, $(n-1)-$point functions. These equations
hold all the information for the fields and their interactions   for any 
number of external legs and to all orders in perturbation theory. 
The recursive content of the Dyson-Schwinger equations for 
QCD has already been introduced  in Ref.\cite{Draggiotis:2002hm} and reviewed 
recently in Ref.\cite{Papadopoulos:2005vg}. To include  the 
electroweak sector,  new vertices for leptons, the vector gauge bosons
as well as for the scalar Higgs boson  must be included. 
Additionally, the recursive equation for (anti)quarks should be rewritten to
express their interaction with the electroweak gauge bosons.
\begin{figure}[!ht]
\begin{center}
\epsfig{file=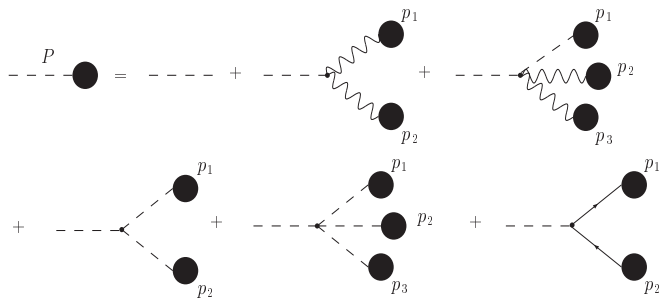,width=90mm,height=40mm}
\end{center}
\caption
{\it  Recursion equations for the Higgs boson.}
\label{Higgs}
\end{figure}

In order to better illustrate this idea let us present as an example 
the  recursive equations for the Higgs boson 
interaction with massive particles only. Let $p_{1},p_{2},\ldots,p_{n}$
represent the external momenta involved in the scattering processs taken 
to be incoming. For a vector field we define a four vector
$b^{\textnormal{\tiny V}}_{\mu}(P)$, which describes any sub-amplitudes 
from which a vector boson $V$ with momentum $P$ can be constructed. 
The momentum $P$ is given as a sum of external particles momenta. 
Accordingly we define a four-dimensional spinor  
$\psi^{\textnormal{\tiny F}}(P)$, which describes any sub-amplitude 
from which a fermion with momentum P can be constructed and by  
$\bar{\psi}^{\textnormal{\tiny F}}(P)$  a four-dimensional antispinor.
Additionaly we have to introduces a scalar $H(P)$  for a Higgs boson.  
The content of
Dyson-Schwinger equations in this case 
can be understood diagrammatically as in Fig.\ref{Higgs}.
The subamplitude with an off-shell Higgs boson momentum $P$
has contributions from three-bosons and  four-bosons vertices plus the fermion 
antifermion vertex. The black blobs denote subamplitudes with the same 
structure. The following general recursive equations can be written down for 
a Higgs boson  with the momentum $P$:
\begin{eqnarray*}
H(P)&&=\sum_{i=1}^{n}\delta(P-p_{i})H(p_{i})\\ 
&&+\sum_{P=p_{1}+p_{2}} ig_{\textnormal{\tiny HVV}}~\Pi_{\textnormal{\tiny H}}
~b^{\textnormal{\tiny V}}_{\mu}(p_{1})b^{{\textnormal{\tiny V}}\mu}
(p_{2})\epsilon(p_{1},p_{2})\\
&&+\sum_{P=p_{1}+p_{2}+p_{3}}ig_{\textnormal{\tiny HHVV}}
~\Pi_{\textnormal{\tiny H}}~H(p_{1})b^{\textnormal{\tiny V}}_{\mu}(p_{2})
b^{{\textnormal{\tiny V}}\mu}(p_{3})
\epsilon(p_{1},p_{2},p_{3})\\
&&+\sum_{P=p_{1}+p_{2}}
ig_{\textnormal{\tiny HFF}}~
\Pi_{\textnormal{\tiny H}}~\bar{\psi}^{\textnormal{\tiny F}}(p_{1})
\psi^{\textnormal{\tiny F}}(p_{2})\epsilon(p_{1},p_{2})\\
&&+\sum_{P=p_{1}+p_{2}} ig_{\textnormal{\tiny HHH}}
~\Pi_{\textnormal{\tiny H}}~
H(p_{1})H(p_{2})\epsilon(p_{1},p_{2})\\
&&+\sum_{P=p_{1}+p_{2}+p_{3}}ig_{\textnormal{\tiny HHHH}}
~\Pi_{\textnormal{\tiny H}}~H(p_{1})H(p_{2})H(p_{3})\epsilon(p_{1},p_{2},p_{3})
\end {eqnarray*}
where the  Higgs boson propagator is given by
\begin{equation}
\Pi_{\textnormal{\tiny H}}=\frac{i}{P^{2}-m^{2}_{\textnormal{\tiny H}}
-i\Gamma_{\textnormal{\tiny H}} m_{\textnormal{\tiny H}}}
\end{equation}
and $\epsilon(p_{1},p_{2},p_{3})=\pm1$ 
is a sign function, which takes into account the 
sign change when two identical fermions are interchanged.  

The scattering amplitude can be calculated by any of the following
relations,
\begin{equation}
{\cal A}(p_1,\ldots,p_n) = \left\{
\begin{array}{ll}
\hat{b}^{\textnormal{\tiny V}}_\mu(P_i) b^{{\textnormal{\tiny V}}\mu}(p_i) 
& - \mbox{ vector bosons}
\\ \\ \hat{H}(P_{i}) H(p_{i})& - \mbox{ Higgs boson}
\\ \\ \hat{\psb^{\textnormal{\tiny F}}}(P_i)
{\psi}^{\textnormal{\tiny F}}(p_i) & - \mbox{ incoming fermion }  
\\\\ \psi^{\textnormal{\tiny F}}(p_i)
\hat{\psi}^{\textnormal{\tiny F}}(P_i) & - \mbox{ outgoing fermion }  \\
\end{array}
\right.
\end{equation}
where
\[ P_i=\sum_{j\not= i}p_j,\]
so that $P_i+p_i=0$. The functions with  hat are given by the 
previous expressions except for the propagator term which is removed 
by the amputation procedure. This is because the 
outgoing momentum $P_{i}$ must be on shell. 
The initial conditions are given by
\begin{eqnarray*}
b^{{\textnormal{\tiny V}}\mu}(p_i)&=&\varepsilon^\mu_\lambda(p_i) ,    ~~~\lambda=\pm 1,0
\nn\\ H(p_{i})&=&1\\
\psi^{\textnormal{\tiny F}}(p_i)&=&\left\{
\begin{array}{ll}
u_\lambda(p_i)
&\mbox{if $E_i\geq0$} \\
v_\lambda(-p_i)
&\mbox{if $E_i\leq0$} \\
\end{array}
\right.
\nn\\
{\psb}^{\textnormal{\tiny F}}(p_i)&=&\left\{
\begin{array}{ll}
\bar{u}_\lambda(p_i)
&\mbox{if $E_i\geq0$} \\
\bar{v}_\lambda(-p_i)
&\mbox{if $E_i\leq0$} \\
\end{array}
\right.
\label{ampl}\end{eqnarray*}
where the explicit form of $\varepsilon^\mu_\lambda,u_\lambda,
v_\lambda,\bar{u}_\lambda,\bar{v}_\lambda$
are given in the Ref.\cite{Kanaki:2000ey}.

In order to actually solve these recursive equations it is convenient to use
a binary representation of the momenta involved \cite{Caravaglios:1995cd}.
For a process with $n$ external particles, to the 
momentum $P^\mu$ defined as
\begin{equation}
P^\mu=\sum_{i=1}^{n}p_i^\mu
\end{equation}
a binary vector $\vec{m}=(m_1,\ldots,m_n)$ can be assigned,
where its components take the values $0$ or $1$, in such a way that
\begin{equation}
P^\mu=\sum_{i=1}^n m_i\;p_i^\mu\;.
\end{equation}
Moreover this binary vector can be uniquely represented by the
integer
\begin{equation}
m=\sum_{i=1}^n 2^{i-1}m_i
\end{equation}
where 
\begin{equation}
1\le m \le 2^{n-1}.
\end{equation}
Therefore all subamplitudes can be labeled accordingly, \ie
\[
\psi^{\textnormal{\tiny F}}(P)\rightarrow\psi^{\textnormal{\tiny F}}(m),
\]
\begin{equation}
\bar{\psi}^{\textnormal{\tiny F}}(P)
\rightarrow\bar{\psi}^{\textnormal{\tiny F}}(m),
\end{equation}
\[
b^{\textnormal{\tiny V}}_\mu(P)\to b^{\textnormal{\tiny V}}_\mu(m),
\]
\[
H(P)\to H(m).
\]
A very convenient ordering of integers in binary representation
relies on the notion of level $l$, defined simply as
\begin{equation}
l=\sum_{i=1}^n m_i\;.
\end{equation}
As it is easily seen all external momenta are of level $1$, whereas
the total amplitude corresponds to the unique level $n$ integer
$2^{n}-1$. This ordering dictates the natural path of the
computation; starting with level-$1$ sub-amplitudes, we compute
the level-$2$ ones using the Dyson-Schwinger equations and so on
up to the level $n$ which is the full amplitude. 

Contrary to original {\tt HELAC} 
\cite{Kanaki:2000ey,Kanaki:2000ms}, the computational part consists of only one
step, where couplings allowed by the Lagrangian defined by 
fusion rules are only
explored. Subsequently, the helicity
configurations are set up. There are two possibilities, either exact summation
over all helicity configurations is performed or Monte Carlo summation 
is applied.  For example for 
a massive gauge boson the second option is achieved by introducing 
the polarization vector  
\begin{equation}
\varepsilon^{\mu}_{\phi}(p)=e^{i\phi}\varepsilon^{\mu}_{+}(p)+
e^{-i\phi}\varepsilon^{\mu}_{-}(p)+\varepsilon^{\mu}_{0}(p),
\end{equation}
where $\phi \in (0,2\pi)$ is a random number. By integrating over 
$\phi$ we can obtain the sum over helicities 
\[
\frac{1}{2\pi}\int_{0}^{2\pi}d\phi ~\varepsilon^{\mu}_{\phi}(p)
(\varepsilon^{\nu}_{\phi}(p))^{*}=\sum_{\lambda=\pm}
\varepsilon_{\lambda}^{\mu}(p)(\varepsilon_{\lambda}^{\nu}(p))^{*}.
\]
The same idea can be applied 
to the helicity of (anti)fermions.

Finally, the color factor is evaluated iteratively. Once again, we
have two options. Either we  
proceed by computing all $3^{n_{q}}\times3^{n_{\bar{q}}}$
color configurations, where the gluon is 
treated as a quark-antiquark pair and $n_{q},n_{\bar{q}}$ is  the number of 
quarks and antiquarks respectively, or Monte Carlo summation is applied.
Only a fraction of all possible $3^{n_{q}}\times3^{n_{\bar{q}}}$  color
configurations gives rise to a non zero amplitude. In the Monte Carlo 
approach for each event we randomly select a non vanishing color 
assignment for the external particles and evaluate the amplitude. 
An overall multiplicative coefficient must be introduced to provide
the correct normalization. The weight of the event is simply
proportional to the $|{\cal M}|^{2}$ multiplied by the number of non zero 
color configurations. Assuming that, on average, all color 
configurations contribute the same amount to the cross section this 
approach is numerically more efficient than summing each event over 
all colors, see Ref.~\cite{new} for further details.

For the spinor wave functions as well as for the Dirac matrices, 
we have chosen the 4-dimensional
chiral representation which results in particularly simple expressions. 
All vertices in the unitary gauge
have been included. Both the fixed width scheme (FWS) and the complex mass 
scheme (CMS) for unstable particles are implemented.  

The computational cost of {\tt HELAC} grows like $\sim 3^n$, which essentially
counts the steps used to solve the recursive equations.
Obviously for large $n$ there is a tremendous saving of computational time,
compared to the $n!$ growth of the Feynman graph approach.


\section{Numerical Results}
As an example the algorithm has been used to compute total cross sections 
for  the production of weak vector bosons and the Higgs boson in association 
with jets. The following Standard Model input parameters have been used
 \cite{Eidelman:2004wy}:
\begin{eqnarray*}
&&m_{W}=80.425 ~\textnormal{GeV},  
~~~~~~~~~~~~~~~~~~\Gamma_{W}=2.124 ~\textnormal{GeV}\\
&&m_{Z}=91.188 ~\textnormal{GeV},  
~~~~~~~~~~~~~~~~~~~\Gamma_{Z}= 2.495 ~\textnormal{GeV}\\
&&G_{\mu}= 1.6637\times10^{-5} ~\textnormal{GeV}^{-5}\\
&&\sin^{2}\theta_{W}=1-m^{2}_{W}/m_{Z}^{2}.
\end {eqnarray*}
The electromagnetic coupling is derived from the Fermi constant $G_{\mu}$
according to
\begin{equation}
\alpha_{\textnormal{\footnotesize em}}=\frac{\sqrt{2}G_{\mu}m^{2}_{W}
\sin^{2}\theta_{W}}{\pi}.
\end{equation}
All results are obtained with a fixed strong coupling constant $\alpha_{s}$ 
calculated at the $m_{Z}$ scale
\begin{equation}
\alpha_{s}(m^{2}_{Z})=0.1187.
\end{equation}
\begin{table}[t!]
\newcommand{\lstrut}{{$\strut\atop\strut$}}
\caption {\em  Results for the total cross section  
for the associated production of 
 Higgs boson (130 GeV) with a $t\bar{t}$ pair, 
$\sigma_{\textnormal {\tiny EXACT}}$ corresponds to summation over
all possible color configurations, while
$\sigma_{\textnormal{\tiny MC}}$ corresponds to
 Monte Carlo summation. }
\label{tab1}
\begin{center}
\begin{tabular}{||l|l|l||}
\hline \hline & & \\
~$\textnormal{Process}$ & ~~~~~~$\sigma_{\textnormal {\tiny EXACT}}$
$\pm$ $\varepsilon$ $\textnormal{(nb)}$ 
& ~~~~~~$\sigma_{\textnormal{\tiny MC}}$
$\pm$ $\varepsilon$ $\textnormal{(nb)}$ \\ & &\\ 
\hline \hline
$gg \rightarrow t\bar{t}H$ & 
(0.2723 $\pm$ 0.0016)$\times 10^{-3}$    &  
(0.2713 $\pm$ 0.0013)$\times 10^{-3}$  \\
\hline  $u\bar{u} \rightarrow  t\bar{t}H$ & 
(0.2758 $\pm$ 0.0017)$\times 10^{-4}$ & 
(0.2739 $\pm$ 0.0011)$\times 10^{-4}$ \\
\hline  $d\bar{d} \rightarrow t\bar{t}H $ & 
(0.1816 $\pm$ 0.0011)$\times 10^{-4}$&  
(0.1811 $\pm$ 0.0007)$\times 10^{-4}$ \\ 
\hline $c\bar{c} \rightarrow t\bar{t}H $ & 
(0.8118 $\pm$ 0.0057)$\times 10^{-6}$ &  
(0.8094 $\pm$ 0.0032)$\times 10^{-6}$   \\
\hline $s\bar{s} \rightarrow t\bar{t}H $ & 
(0.2203 $\pm$ 0.0014)$\times 10^{-5}$   & 
(0.2191 $\pm$ 0.0008)$\times 10^{-5}$ \\
\hline $b\bar{b} \rightarrow t\bar{t}H $ & 
(0.2260 $\pm$ 0.0016)$\times 10^{-6}$  & 
(0.2262 $\pm$ 0.0009)$\times 10^{-6}$ \\
\hline\hline
\end{tabular}
\end{center}
\end{table}
\begin{table}[t!]
\newcommand{\lstrut}{{$\strut\atop\strut$}}
\caption {\em  Results for the total cross section  
for the production of a heavy  
 Higgs boson (200 GeV) via  the vector boson fusion, 
$\sigma_{\textnormal {\tiny EXACT}}$ corresponds to summation over
all possible color configurations, while
$\sigma_{\textnormal{\tiny MC}}$ corresponds to
 Monte Carlo summation. }
\label{tab2}
\begin{center}
\begin{tabular}{||l|l|l||}
\hline \hline  &&\\
~$\textnormal{Process}$ & ~~~~~~$\sigma_{\textnormal {\tiny EXACT}}$
$\pm$ $\varepsilon$ $\textnormal{(nb)}$ 
& ~~~~~~$\sigma_{\textnormal{\tiny MC}}$
$\pm$ $\varepsilon$ $\textnormal{(nb)}$ \\  & & \\
\hline \hline $u\bar{u} \rightarrow u\bar{u}H$ & 
(0.1406 $\pm$ 0.0029)$\times 10^{-5}$    &  
(0.1361 $\pm$ 0.0020)$\times 10^{-5}$ \\
\hline  $u\bar{u} \rightarrow d\bar{d}H$ & 
(0.6699 $\pm$ 0.0088)$\times 10^{-5}$    &  
(0.6596 $\pm$ 0.0081)$\times 10^{-5}$  \\
\hline $ud \rightarrow udH $ &  
(0.2280 $\pm$ 0.0043)$\times 10^{-4}$ &
(0.2222 $\pm$ 0.0021)$\times 10^{-4}$\\
\hline $\bar{u}d \rightarrow \bar{u}dH $ & 
(0.1241 $\pm$ 0.0027)$\times 10^{-5}$  & 
(0.1258 $\pm$ 0.0025)$\times 10^{-5}$\\
\hline  $dd \rightarrow ddH $ & 
(0.3404 $\pm$ 0.0046)$\times 10^{-5}$  & 
(0.3477 $\pm$ 0.0032)$\times 10^{-5}$\\
\hline $uu \rightarrow uuH $ & 
(0.5132 $\pm$ 0.0081)$\times 10^{-5}$  & 
(0.5178 $\pm$ 0.0060)$\times 10^{-5}$\\
\hline\hline
\end{tabular}
\end{center}
\end{table}

\begin{figure}[!ht]
\begin{center}
\epsfig{file=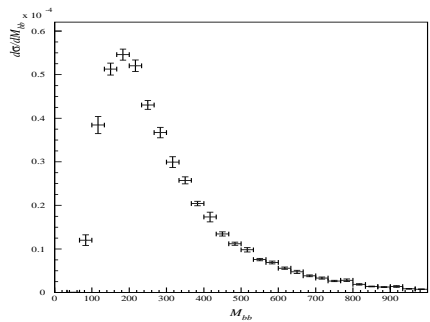,width=62mm,height=45mm}
\end{center}
\caption{\em Invariant mass distribution of the $bb$ system in 
$gg\rightarrow W^{+}W^{-}b\bar{b}b\bar{b}$ process.}
\label{inv1}
\end{figure}
\begin{figure}[!ht]
\begin{center}
\epsfig{file=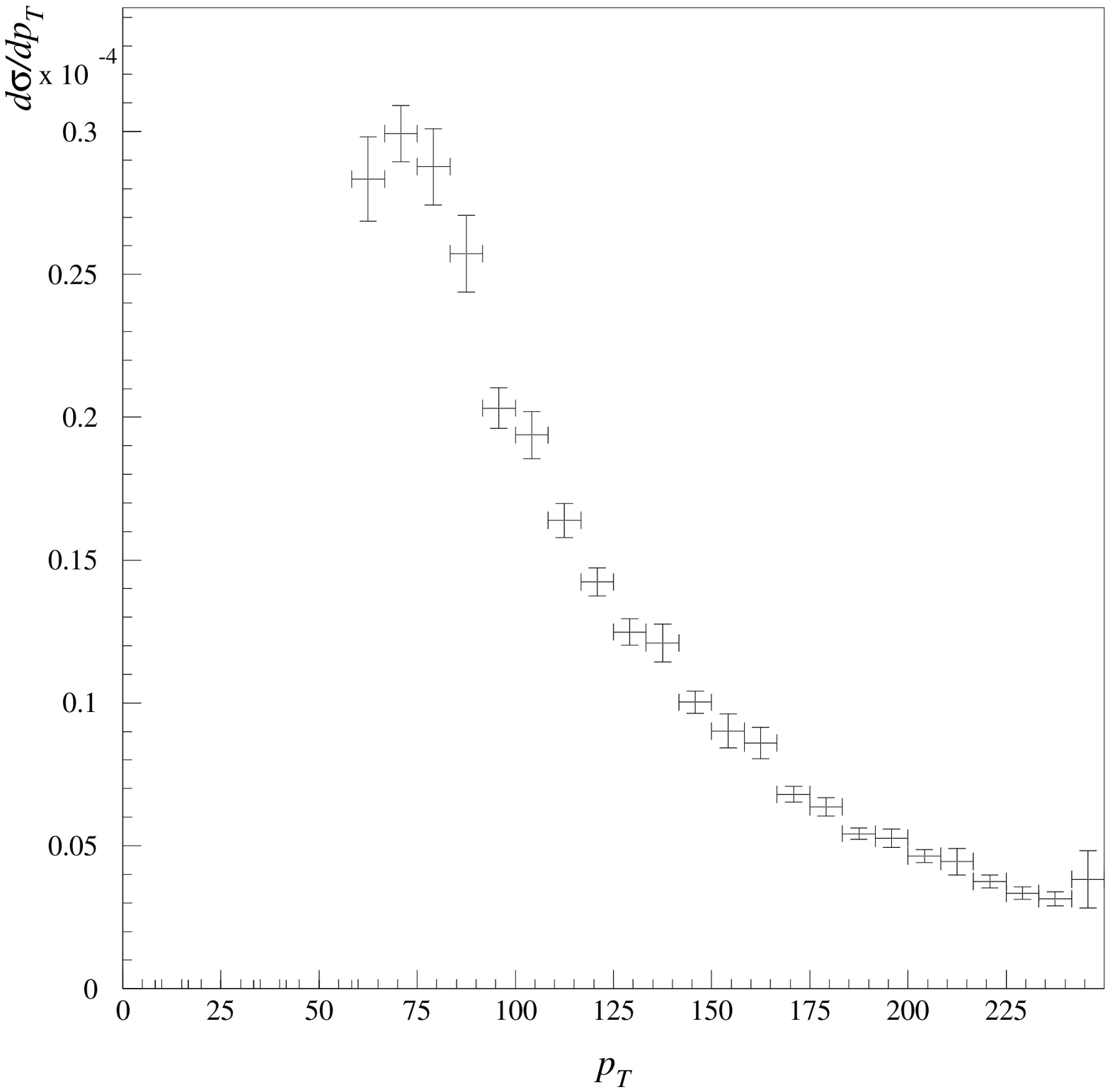,width=62mm,height=45mm}
\epsfig{file=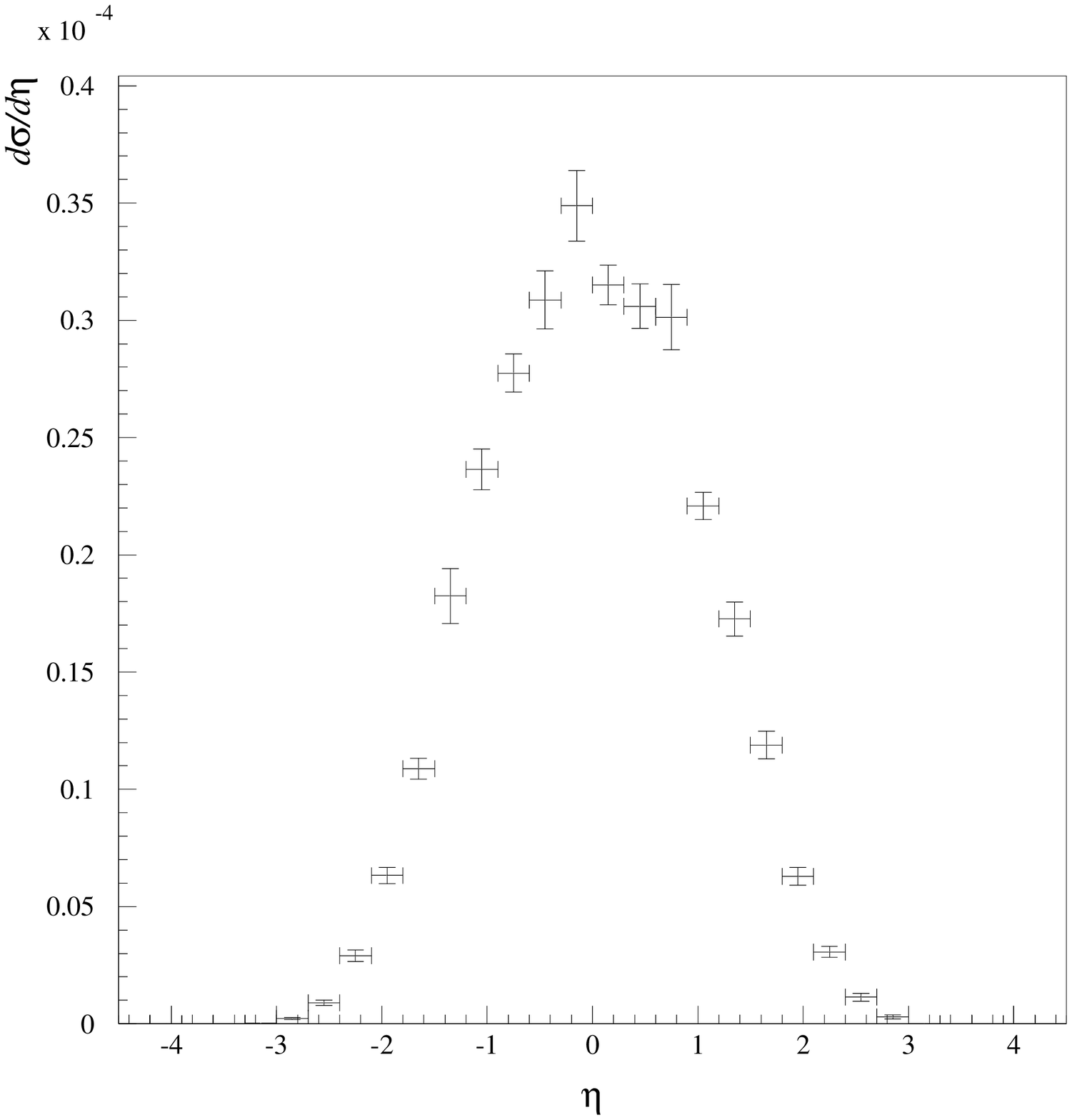,width=62mm,height=45mm}
\end{center}
\caption{\em Transverse momentum distribution (left panel) 
and  rapidity distribution (right panel) of  $b$-jet in 
$gg\rightarrow W^{+}W^{-}b\bar{b}b\bar{b}$ process.}
\label{pt-eta-b}
\end{figure}
The mass of an intermediate and a heavy Higgs boson and associated 
Standard Model tree level widths are assumed to be:
\begin{eqnarray*}
&&m_{H}=130 ~\textnormal{GeV},  
~~~~~~~~~~~~~~~~~~~\Gamma_{H}= 0.005 ~\textnormal{GeV}\\
&&m_{H}=200 ~\textnormal{GeV},
~~~~~~~~~~~~~~~~~~~\Gamma_{H}= 1 ~\textnormal{GeV}\\
\end {eqnarray*}
For the massive fermions the following masses have been applied:
\begin{eqnarray*}
&&m_{u}=4 ~\textnormal{MeV}, ~~~~~~~~~~~~~~~~~~~m_{d}=8 ~\textnormal{MeV}\\
&&m_{s}=130 ~\textnormal{MeV}, ~~~~~~~~~~~~~~~~~m_{c}=1.35 ~\textnormal{GeV}\\
&&m_{b}=4.4 ~\textnormal{GeV},\\
&&m_{t}= 174.3 ~\textnormal{GeV}, ~~~~~~~~~~~~~~~\Gamma_{t}=1.56 ~\textnormal{GeV}\\
\end {eqnarray*}
The mixing of the quark generations is neglected.

The CMS energy was chosen $\sqrt{s}=14$ $\textnormal{TeV}$.
The following cuts were used to stay away from soft and collinear 
divergencies in the part of the phase space intergated over: 
\begin{equation}
p_{T_{i}} > 60 ~\textnormal{GeV}, ~~~~~~ |y_{i}|<2.5, ~~~~~~ \Delta R > 1.0
\end{equation}
where 
\begin{equation}
p_{T_{i}}=\sqrt{p_{x_{i}}^{2}+p_{y_{i}}^{2}}, ~~~~~~~~ 
y_{i}=\frac{1}{2}\ln \left( \frac{ E_{i}+p_{z_{i}} }{ E_{i}-p_{z_{i}} }\right)
\end{equation}
are the transverse momentum and 
rapidity of the $i$-jet respectively. Additionaly $\Delta R$ is a radius
of the cone of the jet defined as 
\[
\Delta R=\sqrt{(\Phi_{i}-\Phi_{j})^{2}+(y_{i}-y_{j})^2}
\]
where $\Delta \Phi_{ij}=\Phi_{i}-\Phi_{j}$ 
\[ 
\Delta \Phi_{ij}=\arccos \left( \frac{p_{x_{i}}p_{x_{j}}+p_{y_{i}}p_{y_{j}}}
{p_{T_{i}} p_{T_{j}} } \right).
\]

\begin{figure}[!ht]
\begin{center}
\epsfig{file=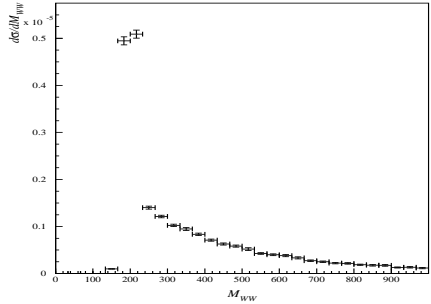,width=62mm,height=45mm}
\end{center}
\caption{\em Invariant mass distribution of the $W^{+}W^{-}$ system in 
$u\bar{u}\rightarrow W^{+}W^{-}d\bar{d}$ process.}
\label{inv2}
\end{figure}
\begin{figure}[!ht]
\begin{center}
\epsfig{file=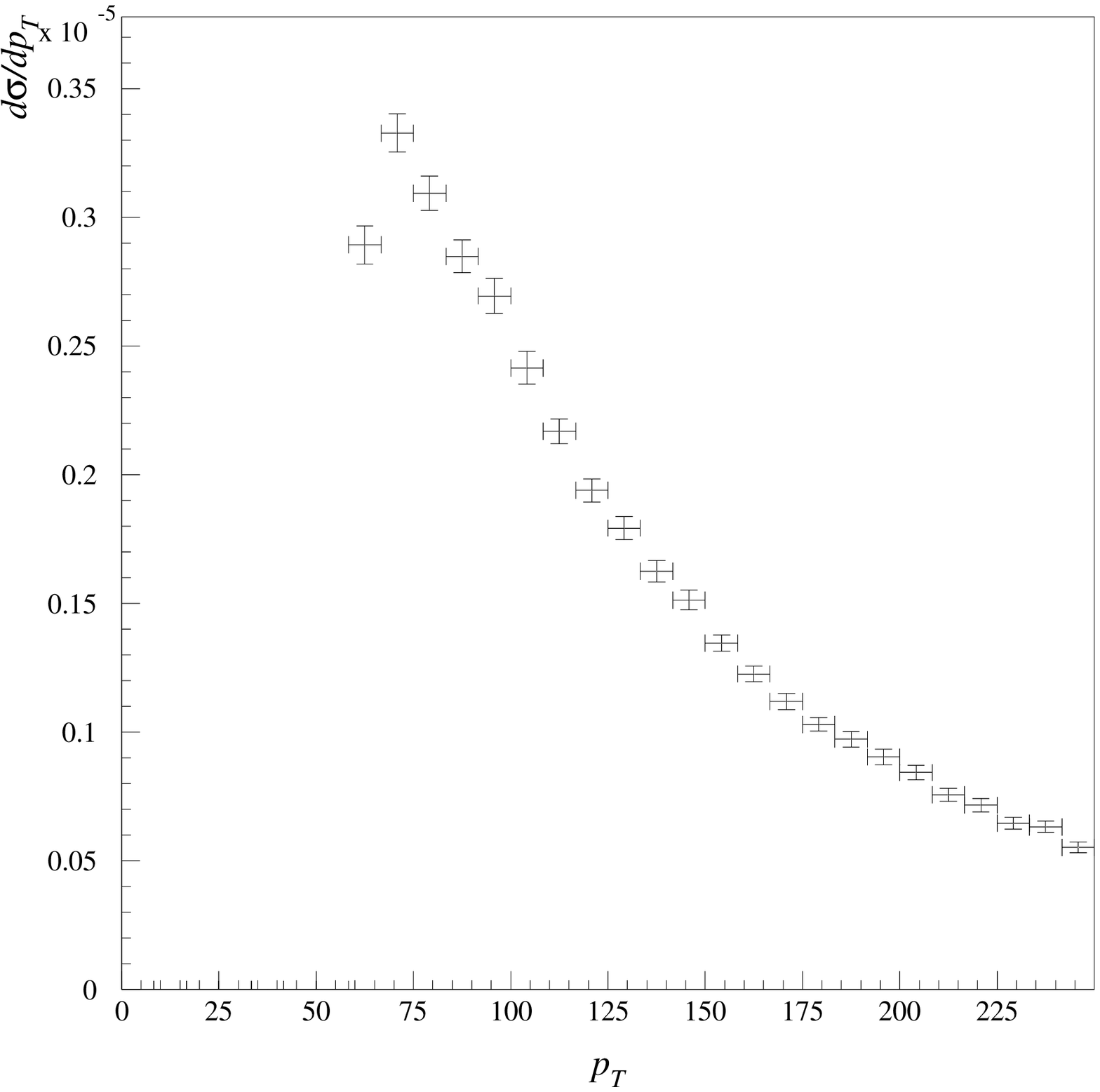,width=62mm,height=45mm}
\epsfig{file=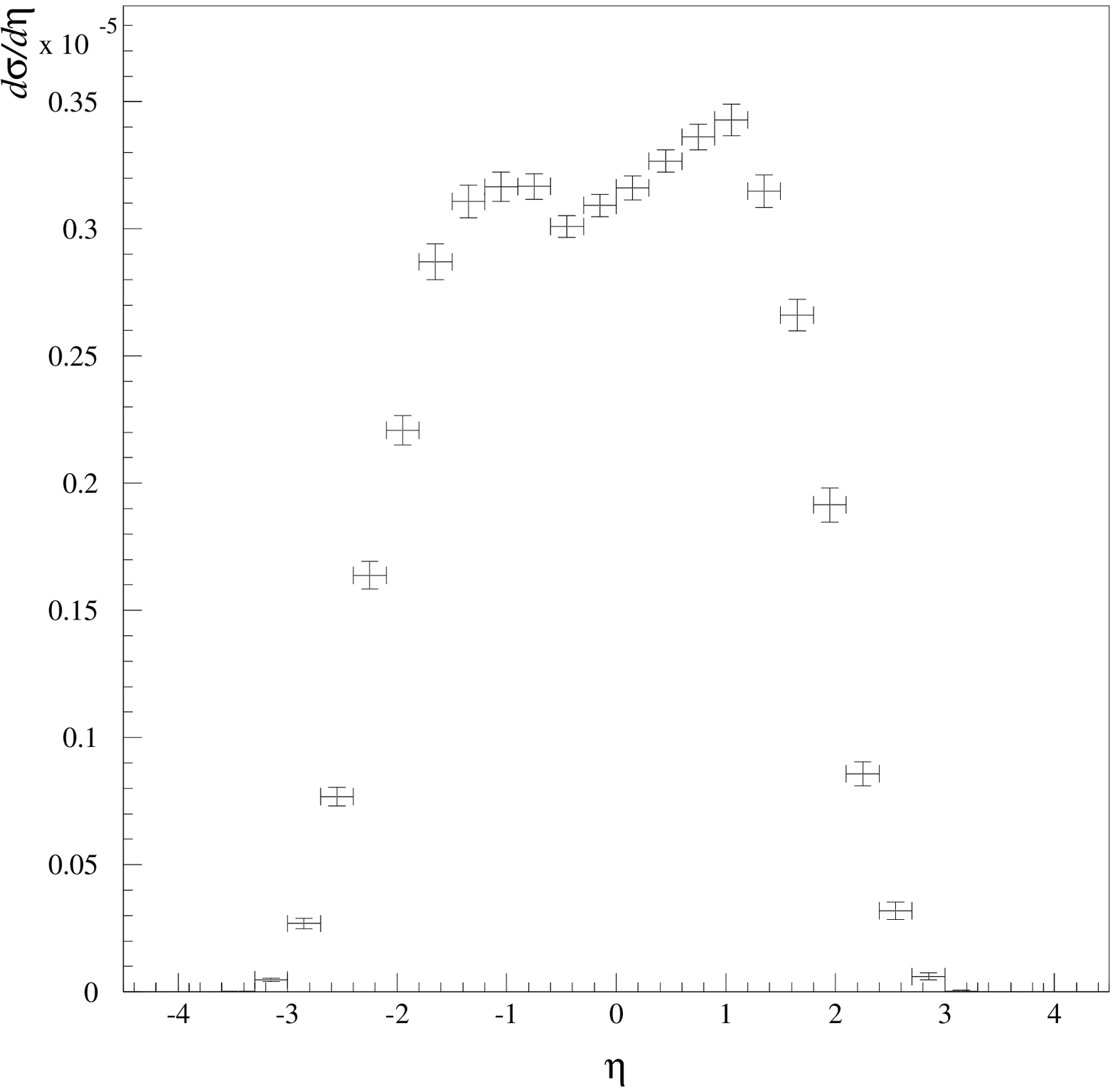,width=62mm,height=45mm}
\end{center}
\caption{\em Transverse momentum distribution (left panel) 
and  rapidity distribution (right panel) of d-jet  in 
$u\bar{u}\rightarrow W^{+}W^{-}d\bar{d}$ process.}
\label{pt-eta-d}
\end{figure}
\begin{figure}[!ht]
\begin{center}
\epsfig{file=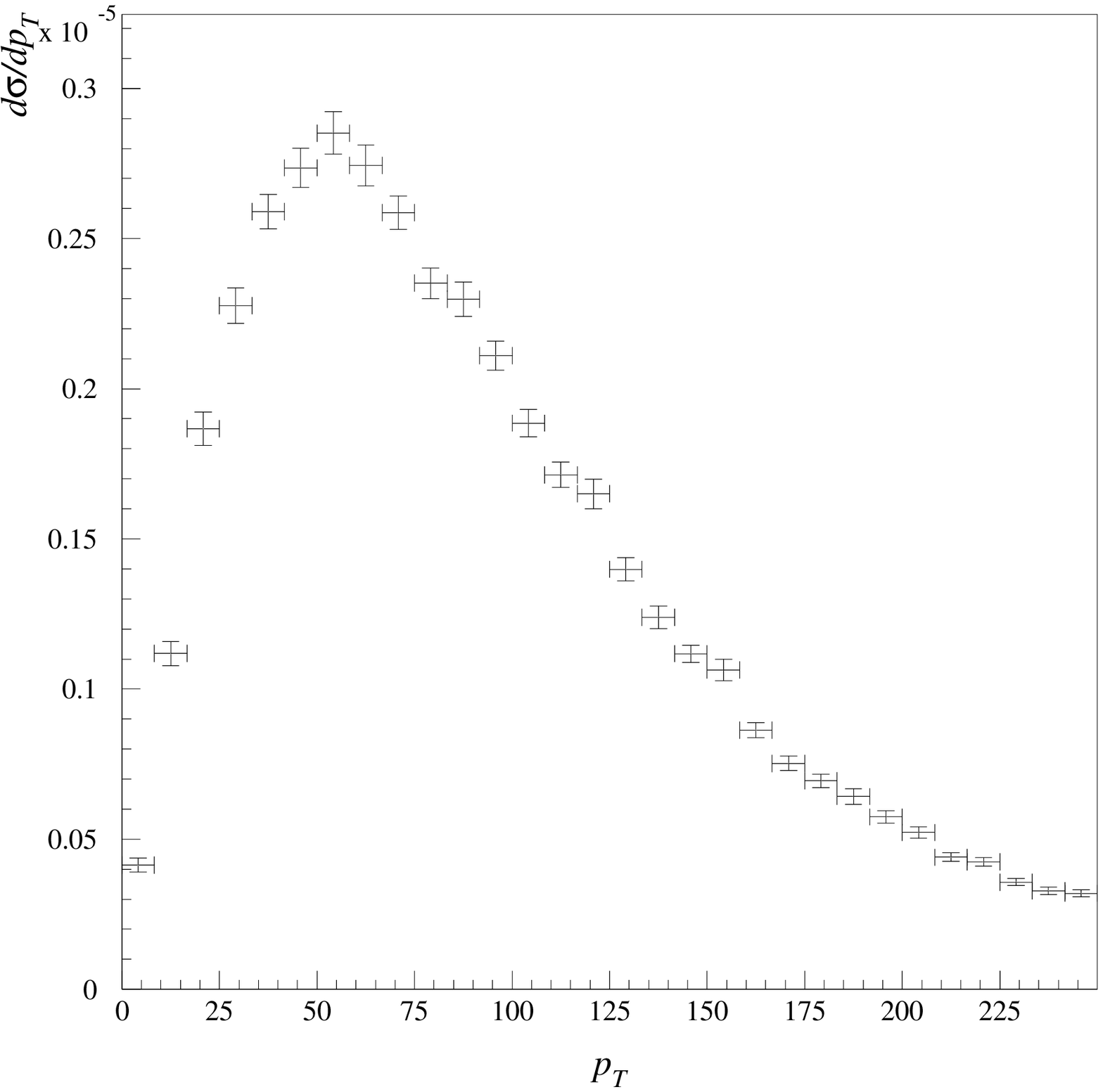,width=62mm,height=45mm}
\epsfig{file=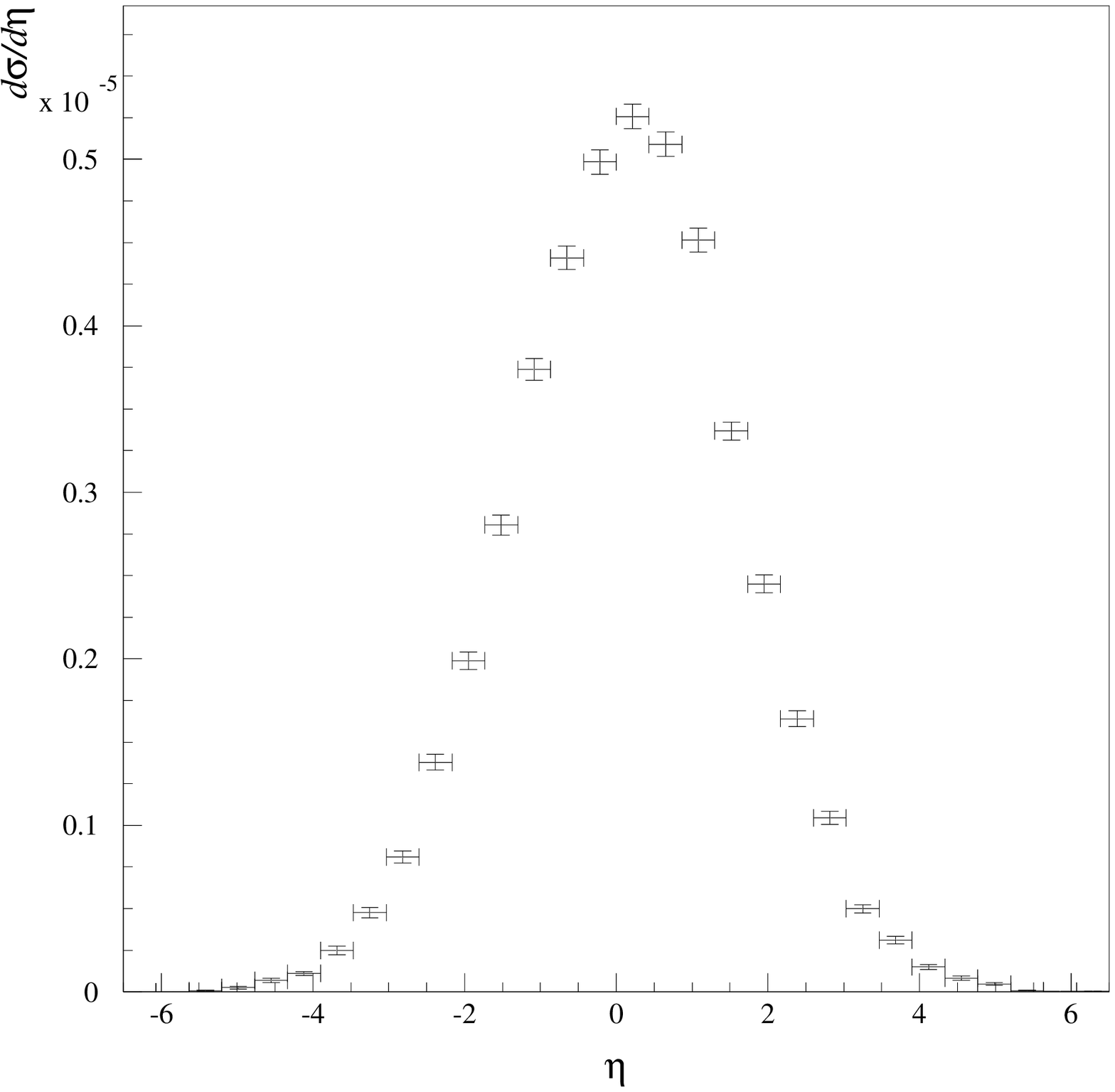,width=62mm,height=45mm}
\end{center}
\caption{\em Transverse momentum
 distribution (left panel) and rapidity distribution
(right panel) of $W$ in 
$u\bar{u}\rightarrow W^{+}W^{-}d\bar{d}$ process. }
\label{pt-eta-w}
\end{figure}

There are several parameterizations for the parton structure functions, we 
used  {\tt CTEQ6 PDF}'s parametrization \cite{Pumplin:2002vw,Stump:2003yu}. 
For the phase space generation we used two different generators.
The first one is {\tt PHEGAS} \cite{Papadopoulos:2000tt}, which
automatically constructs mappings of all possible peaking structures of a given 
scattering process. Self-adaptive procedures like multi-channel 
optimization \cite{Kleiss:1994qy} is additionally applied exhibiting
 high efficiency. The
Second one is  a flat phase-space generator {\tt RAMBO} \cite{Kleiss:1985gy}.

In the Table \ref{tab1}. results for the total cross section  
for the associated production of the
 Higgs boson of $m_{H}=130$ GeV mass, 
with a $t\bar{t}$ pair are presented.  The $t\bar{t}H$
production channel is one of the most promising reactions to study both 
the top quark and the Higgs boson at the LHC,  in the second case especially for
the $b\bar{b}$ decay channel of the Higgs boson 
\cite{Beneke:2000hk,Djouadi:2005gi}. As we can see from the Table \ref{tab1}. 
the $gg\rightarrow t\bar{t}H$ process dominates 
due to the enhanced gluon structure function. The final state of this channel consists
of $W$ bosons and four $b-$jet, two from the decay of the top quarks
and two from the decay of the Higgs boson. The main background process 
is $gg\rightarrow W^{+}W^{-}b\bar{b}b\bar{b}$ with the contributions from 
all intermediate states. In Fig.\ref{inv1} we present the 
invariant mass distribution for  the $b\bar{b}$ system.   
Transverse momentum and rapidity distributions of
$b$-jet for the background  process are shown in Fig. \ref{pt-eta-b}.

A more powerful channel for higher Higgs boson masses is the vector 
boson fusion $qq\rightarrow V^{*}V^{*}\rightarrow qqH$ with 
$H\rightarrow W^{+}W^{-}$ decay.
In the Table \ref{tab2}. results for the total cross section  for some 
production processes of a heavy  Higgs boson of $m_{H}=200$ GeV mass  
via vector boson fusion  are presented.  The main background 
processes consist of $qq \rightarrow  W^{+}W^{-}qq$ channels.  
As an example the distribution of the invariant mass of the $W^{+}W^{-}$ system
from the $u\bar{u} \rightarrow  W^{+}W^{-}d \bar{d}$ process 
is presented in Fig.\ref{inv1}. Transverse momentum and rapidity  
distributions  of $d$-jet  and $W$ 
are also shown in Fig.\ref{pt-eta-d} and Fig.\ref{pt-eta-w}.

\section*{Summary}
An efficient tool for  automatic computation of helicity
amplitudes and cross sections for multi-particle final states in
the Standard Model has been presented. Matrix elements and cross sections 
are calculated iteratively by Dyson-Schwinger equations. We are free 
from the task of computing all Feynman diagrams for a given process, which 
can become impossible  for the large number of particles involved. The 
computationally expensive procedures of summing over color and helicity 
configurations have been replaced by Monte Carlo summation. At this stage,
 the code is able to compute scattering matrix elements and cross sections 
for hard processes. In the next step we plan 
to make calculations of fully hadronic final states
in $p\bar{p}$ and $pp$ collisions possible. In particular, we wish to include the 
the emission of secondary partons 
and translate the emerging partons into primordial hadrons \eg 
by interfacing this package with codes like 
{\tt PYTHIA}\cite{Sjostrand:2000wi} or {\tt HERWIG}\cite{Corcella:2000bw}.
This kind of multipurpose Monte Carlo generator will be of great
interest in the study of the TeVatron, LHC or $e^{+}e^{-}$ 
Linear Collider data.

\section*{Acknowledgments}
Work supported by the Polish State Committee for Scientific
Research Grants number 1 P03B 009 27 for years 2004-2005
(M.W.).  In addition,
M.W. acknowledges the Maria Curie Fellowship granted by the
the European Community in the framework of the Human Potential Programme
under contract HPMD-CT-2001-00105
({\it ``Multi-particle production and higher order correction''}).
The Greece-Poland bilateral agreement {\it ``Advanced computer 
techniques for theoretical calculations and development of simulation 
programs for high energy physics experiments''} is also acknowledged.


\providecommand{\href}[2]{#2}

\end{document}